\documentclass[10pt,a4paper]{article}

\usepackage[latin1]{inputenc} 
\usepackage[dvips]{graphicx} 
\usepackage{a4wide}
\usepackage{latexsym}
\usepackage{amsmath}
\usepackage{amssymb}
\usepackage{dsfont}
\usepackage{color}

\title{Electromagnetic Properties for Arbitrary Spin Particles:\\
Part 2 $-$ Natural Moments and Transverse Charge Densities}

\author{Cédric Lorcé\\ \small{\emph{Institut für Kernphysik, Johannes Gutenberg-Universität, D-55099 Mainz, Germany}}\\
\small{\emph{E-mail: lorce@kph.uni-mainz.de}}}
\date{}

\newcommand{\ud}{\mathrm{d}}

\begin{document}

\maketitle

\begin{center}
\begin{minipage}[t]{15cm}
\small{In a set of two papers, we propose to study an old-standing problem, namely the electromagnetic interaction for particles of arbitrary spin. Based on the assumption that light-cone helicity at tree level and $Q^2=0$ should be conserved non-trivially by the electromagnetic interaction, we are able to derive \emph{all} the natural electromagnetic moments for a pointlike particle of \emph{any} spin. In this second paper, we give explicit expressions for the light-cone helicity amplitudes in terms of covariant vertex functions, leading to the natural electromagnetic moments at $Q^2=0$. As an application of our results, we generalize the discussion of quark transverse charge densities to particles with arbitrary spin.}
\end{minipage}
\end{center}

\section{Introduction}

As explained in Part 1 of this work \cite{partim1}, the question about the natural values of particle electromagnetic moments is an important one for many reasons. Previous studies have mainly focused on the gyromagnetic factor $g_j$ of a spin-$j$ elementary particle \cite{gyro,indep}, which at tree level is required to take on the value $g_j=2$ to maintain unitarity. More recent studies proposed also some discussion on higher multipoles \cite{higher,higherbis}, such as the electric quadrupole moment.

In a set of two papers, we present our results concerning the electromagnetic interaction for particles with arbitrary spin. In the first paper, we discussed the electromagnetic vertex for a spin-$j$ particle, defined corresponding covariant vertex functions (or form factors) and performed a multipole decomposition of the current. In the present paper, we firstly study in Section \ref{LCHA} the light-cone helicity amplitudes for the electromagnetic vertex of a particle of any spin. Using the assumption that light-cone helicity is conserved non-trivially at tree level and at the real photon point ($Q^2=0$) by the electromagnetic interaction for elementary particles, we obtain in Section \ref{NEM} the natural values of covariant vertex functions, multipole form factors and therefore electromagnetic moments. As an application of our results, we generalize in Section \ref{TS} to arbitrary spin particles the study of quark transverse charge densities, recently discussed in the literature for spin $1/2$ \cite{spinhalf}, spin $1$ \cite{spinone} and spin $3/2$ \cite{paper}.

\section{Light-Cone Helicity Amplitudes}\label{LCHA}

We consider here the light-cone helicity amplitudes of the $+$ component\footnote{We define $x^{\pm}\equiv x^0\pm x^3$.} of the current $J^\mu_\textrm{EM}$. We will work in the usual Drell-Yan-West (DYW) frame $q^+=0$ \cite{DYW}, where $q$ is the four-momentum of the photon. We can furthermore choose a frame where the transverse momenta of the initial and final particles are opposite. We will write such light-cone helicity amplitude in the form
\begin{equation}
A_{\lambda',\lambda}(Q^2)\equiv\frac{e^{i(\lambda'-\lambda)\phi_q}}{2p^+}\,\langle
p^+,\frac{\vec{q}_\perp}{2},\lambda'|J^+_\textrm{EM}(0)|p^+,-\frac{\vec{q}_\perp}{2},\lambda\rangle,
\end{equation}
where $Q^2\equiv-q^2=\vec{q}_\perp^{\,2}$, with $\vec{q}_\perp=Q\left(\cos\phi_q\,\hat e_x+\sin\phi_q\,\hat e_y\right)$, and $\lambda,\lambda'$ are the light-cone helicities of the initial and final particles, respectively.

A spin-$j$ particle has $2j+1$ possible polarization states. This means that there are in principle $(2j+1)^2$ helicity amplitudes. However, as already mentioned in Part 1 \cite{partim1}, one needs only $2j+1$ covariant vertex functions to write the on-shell vertex function \cite{ind}. This means that out of the $(2j+1)^2$ helicity amplitudes, only $2j+1$ are in fact \emph{independent}. One needs therefore $2j(2j+1)$ constraints. It appeared that these constraints arise due to, on the one hand, discrete space-time symmetries, and on the other hand, angular momentum conservation.

\subsection{Light-Cone Discrete Symmetries}

Obviously, parity and time-reversal discrete symmetries
are not compatible with the DYW frame $q^+=0$. However,
relevant light-cone parity and time-reversal operators can be defined
by compounding the usual parity and time-reversal operators with
$\pi$-rotation about the $y$-axis, choosing the $x$-axis so that all momenta lie in the $x$-$z$ plane
\cite{LCdiscrete}.

This results in the parity relation for light-cone helicity amplitudes and identical
particles, given by
\begin{equation}\label{parity}
A_{-\lambda',-\lambda}(Q^2)=(-1)^{\lambda'-\lambda}\,A_{\lambda',\lambda}(Q^2),
\end{equation}
while the time-reversal relation reads
\begin{equation}\label{time-reversal}
A_{\lambda,\lambda'}(Q^2)=(-1)^{\lambda'-\lambda}\,A_{\lambda',\lambda}(Q^2).
\end{equation}
From these relations, one easily deduces that the number of pertinent
amplitudes is reduced to
$\left(j+\frac{1}{2}\right)\left(j+\frac{3}{2}\right)$ and
$\left(j+1\right)^2$ for half-integer and integer spin,
respectively.

\subsection{Light-Cone Angular Conditions}

We know that at the end there should only be $2j+1$ independent light-cone amplitudes. The remaining contraints to be imposed are provided by considerations of angular momentum conservation. Such relations are known in the literature as \emph{angular conditions}. The link between these relations and angular momentum conservation can be made transparent in the Breit frame. This has been discussed in \cite{Helicity,Angcond}. The number of angular conditions is obviously $j^2-\frac{1}{4}$ for half-integer spin, and $j^2$ for integer spin.

Since we will work with the explicit expression for helicity amplitudes in terms of covariant vertex functions, we automatically satisfy angular momentum conservation. There are therefore two ways to obtain the angular conditions. One is to directly impose angular momentum conservation at the level of amplitudes, referring to a specific frame where this condition is simple (Breit frame) and then performing a transformation to the light-cone frame. The other possibility is to directly work with the explicit decomposition in terms of covariant vertex functions. Light-cone helicity amplitudes then appear as linear combinations of these covariant vertex functions, and only a subset of them constitutes linearly independent combinations. As we will show later, one can choose for example the set $\left\{A_{j,j-m}|m=0,\cdots,2j\right\}$ as the independent amplitudes. Any other light-cone helicity amplitude can therefore be written as a linear combination of the elements of this set, arising either from a discrete light-cone symmetry relation (only one element of the independent amplitudes is needed) or from an angular condition (many elements of the independent amplitudes are needed).

\subsection{Helicity Amplitudes and Covariant Vertex Functions}

In Part 1 \cite{partim1}, we have written the on-shell vertex function explicitly in terms of covariant vertex functions, with the assumption that parity and time-reversal symmetries are respected. To obtain the light-cone helicity amplitudes precisely in terms of these covariant vertex functions, we just have to consider the $+$ component of the current $J^+$ and the proper expressions for the light-cone spinors
\begin{equation}
u(p,+)=\frac{1}{\sqrt{2p^+}}\left(\begin{array}{c}p^++M\\ p_R\\ p^+-M\\ p_R\end{array}\right),\qquad u(p,-)=\frac{1}{\sqrt{2p^+}}\left(\begin{array}{c}-p_L\\ p^++M\\ p_L\\ -p^++M\end{array}\right),
\end{equation}
where
\begin{equation}
p_{R,L}\equiv p_x\pm ip_y,
\end{equation}
and the light-cone polarization four-vectors
\begin{equation}
\begin{split}
\varepsilon^\mu(p,+)&=\,-\frac{1}{\sqrt{2}}\left(0,\frac{2p_R}{p^+},\hat{e}_R\right),\\
\varepsilon^\mu(p,0)&=\,\frac{1}{M}\left(p^+,\frac{p_Rp_L-M^2}{p^+},\frac{p_R\,\hat{e}_L+p_L\,\hat{e}_R}{2}\right),\\
\varepsilon^\mu(p,-)&=\,\frac{1}{\sqrt{2}}\left(0,\frac{2p_L}{p^+},\hat{e}_L\right),
\end{split}
\end{equation}
where
\begin{equation}
\hat{e}_{R,L}\equiv \hat{e}_x\pm i\hat{e}_y.
\end{equation}

We will work in the symmetric light-cone frame which is the DYW frame $q^+=0$ where the light-cone energy $p^-$ is conserved
\begin{equation}
\begin{split}
&q^\mu=(0,0,\vec{q}_\perp),\\
&p^\mu=(p^+,\frac{M^2\left(1+\tau\right)}{p^+},-\frac{\vec{q}_\perp}{2})\\
&p'^\mu=(p^+,\frac{M^2\left(1+\tau\right)}{p^+},\frac{\vec{q}_\perp}{2}),
\end{split}
\end{equation}
where $p$ and $p'$ are the four-momenta of the incoming and outgoing particle, respectively and $\tau\equiv Q^2/4M^2$.

Let us consider only the set of amplitudes $\left\{A_{j,j-m}|m=0,\cdots,2j\right\}$, separately for integer and half-integer spin cases. The binomial function $C_n^k$ used in the following is defined as
\begin{equation}
C_n^k=\left(\begin{array}{c}n\\ k\end{array}\right)\equiv\left\{\begin{array}{cl} \frac{n!}{k!\,(n-k)!},&n\geq k\geq 0\\ 0,& \textrm{otherwise}\end{array}\right..
\end{equation}

\subsubsection{Arbitrary Integer Spin Case}

The polarization tensor of the outgoing spin-$j$ particle is very simple \cite{partim1}
\begin{displaymath}
\varepsilon_{\alpha_1\cdots\alpha_j}^*(p',j)=\prod_{l=1}^j\varepsilon^*_{\alpha_l}(p',+).
\end{displaymath}
Thanks to the following relations in the DYW frame
\begin{equation}\label{LCid}
\begin{split}
\varepsilon^*(p',+)\cdot\varepsilon(p,0)&=\,\sqrt{2\tau}\,e^{-i\phi_q}\left[\varepsilon^*(p',+)\cdot\varepsilon(p,+)\right],\\
\varepsilon^*(p',+)\cdot\varepsilon(p,-)&=\,0,\\
\varepsilon(p,0)\cdot q&=\,\sqrt{2\tau}\,e^{-i\phi_q}\left[\varepsilon(p,+)\cdot q\right],\\
\varepsilon(p,-)\cdot q&=\,-e^{-2i\phi_q}\left[\varepsilon(p,+)\cdot q\right],
\end{split}
\end{equation}
we can rewrite all structures in terms of $\varepsilon^*(p',+)\cdot\varepsilon(p,+)$ and $\left[\varepsilon^*(p',+)\cdot q\right]\left[\varepsilon(p,+)\cdot q\right]$ only. Moreover, since we have
\begin{equation}\label{LCid2}
\begin{split}
\varepsilon^*(p',+)\cdot\varepsilon(p,+)&=\,-1,\\
-\frac{\left[\varepsilon^*(p',+)\cdot q\right]\left[\varepsilon(p,+)\cdot q\right]}{2M^2}&=\,-\tau,
\end{split}
\end{equation}
and using the polarization tensor for the incoming particle with the current decomposition Eq. (5) of Part 1 \cite{partim1}, we finally obtain the light-cone helicity amplitudes in terms of covariant vertex functions $F_k(q^2)$
\begin{equation}\label{LCint}
A_{j,j-m}(Q^2)=\frac{\left(4\tau\right)^{m/2}}{\sqrt{C_{2j}^m}}\sum_{k=0}^j\sum_{t=0}^{\min\{k,m/2\}}\frac{\left(-1\right)^t}{2^{2t}}\,C_k^t\,\tau^{k-t}\left[C_{j-t}^{m-2t}\,F_{2k+1}(Q^2)-\frac{1-\delta_{k,j}}{2}\,C_{j-t-1}^{m-2t-1}\,F_{2k+2}(Q^2)\right].
\end{equation}

\subsubsection{Arbitrary Half-Integer Spin Case}

The spin-tensor of the outgoing spin-$j$ particle is very simple \cite{partim1}
\begin{displaymath}
\bar u_{\alpha_1\cdots\alpha_n}(p',j)=\bar u(p',+)\prod_{l=1}^n\varepsilon^*_{\alpha_l}(p',+),
\end{displaymath}
where $j=n+\frac{1}{2}$.
Thanks to \eqref{LCid}, \eqref{LCid2} and the following relations in the DYW frame
\begin{equation}\label{LCid3}
\begin{split}
\bar u(p',+)\,\gamma^+\,u(p,+)&=\,2p^+,\\
\bar u(p',+)\,\gamma^+\,u(p,-)&=\,0,\\
\bar u(p',+)\,\frac{i\sigma^{+\nu}q_\nu}{2M}\,u(p,+)&=\,0,\\
\bar u(p',+)\,\frac{i\sigma^{+\nu}q_\nu}{2M}\,u(p,-)&=\,-\sqrt{\tau}\,e^{-i\phi_q}\,2p^+,
\end{split}
\end{equation}
and using the polarization tensor for the incoming particle with the current decomposition Eq. (8) of Part 1 \cite{partim1}, we finally obtain the light-cone helicity amplitudes in terms of covariant vertex functions $F_k(q^2)$
\begin{equation}\label{LChalfint}
A_{j,j-m}(Q^2)=\frac{\left(4\tau\right)^{m/2}}{\sqrt{C_{2j}^m}}\sum_{k=0}^n\sum_{t=0}^{\min\{k,m/2\}}\frac{\left(-1\right)^t}{2^{2t}}\,C_k^t\,\tau^{k-t}\left[C_{n-t}^{m-2t}\,F_{2k+1}(Q^2)-\frac{1}{2}\,C_{n-t}^{m-2t-1}\,F_{2k+2}(Q^2)\right].
\end{equation}

\section{Natural Electromagnetic Moments}\label{NEM}

First of all, let us notice that the light-cone helicity amplitudes for integer \eqref{LCint} and half-integer spin \eqref{LChalfint} can in fact be written in one unique formula
\begin{equation}\label{LC}
A_{j,j-m}(Q^2)=\frac{\left(4\tau\right)^{m/2}}{\sqrt{C_{2j}^m}}\sum_{k=0}^{[j]}\sum_{t=0}^{\min\{k,m/2\}}\frac{\left(-1\right)^t}{2^{2t}}\,C_k^t\,\tau^{k-t}\left[C_{[j]-t}^{m-2t}\,F_{2k+1}(Q^2)-\frac{1-\delta_{k,j}}{2}\,C_{[j-\frac{1}{2}]-t}^{m-2t-1}\,F_{2k+2}(Q^2)\right],
\end{equation}
where $[x]$ means the largest integer $k$ such that $k\leq x$.

Since we are interested in electromagnetic moments, let us consider the limit $Q^2=0$. Obviously, only the case $m=0$ does not vanish 
\begin{equation}
A_{j,j}(0)=F_1(0)\equiv Z,
\end{equation}
where $Z$ stands for the particle charge in units of $e$, so that an electron has $Z=-1$. In other words, the helicity-conserving amplitude at $Q^2=0$ just gives the electric charge of the particle.

Due to the factor $\tau^{m/2}$ in \eqref{LC}, a light-cone amplitude involving $m$ units of helicity flip behaves at least as $Q^m$ when $Q^2\to 0$. Let us define new light-cone helicity amplitudes where this trivial $Q^2$-dependence is removed
\begin{equation}
G_{j,j-m}(Q^2)\equiv Q^{-m}\,A_{j,j-m}(Q^2).
\end{equation} 
At $Q^2=0$, these amplitudes read
\begin{equation}\label{LCrem}
G_{j,j-m}(0)=\frac{1}{M^m\sqrt{C_{2j}^m}}\,\sum_{k=0}^m\frac{\left(-1\right)^{\left[\frac{k+1}{2}\right]}}{2^k}\,C_{[j-k/2]}^{m-k}\,F_{k+1}(0).
\end{equation}

We can now show that the set $\left\{A_{j,j-m}|m=0,\cdots,2j\right\}$ we chose consists of independent amplitudes only. If the $A_{j,j-m}(Q^2)$ were dependent, this would also be the case for the $A_{j,j-m}(0)$. This means that if we show that the $A_{j,j-m}(0)$ are independent, this will also be true for the $A_{j,j-m}(Q^2)$. Note now that in \eqref{LCrem}, the highest possible covariant vertex function involved is $F_{m+1}(0)$. Consequently, the $A_{j,j-m}(0)$ are independent as well as the $A_{j,j-m}(Q^2)$.

Moreover, thanks to the relations in the DYW frame Eqs. \eqref{LCid}-\eqref{LCid3}, one can see that we have in fact
\begin{equation}\label{consequence}
G_{j-k,j-k-m}(0)=G_{j,j-m}(0),\qquad\forall\, k\in [0,2j-m].
\end{equation}

\subsection{Helicity Conservation}

In order to derive the natural electromagnetic moments of any particle, we need an assumption concerning the electromagnetic interaction. We propose to \emph{assume} that, at tree level and $Q^2=0$, the light-cone helicity of any elementary particle is \emph{non-trivially} conserved. In other words, we assume that
\begin{equation}\label{assumption}
G_{j,j-m}(0)=\delta_{m,0}\,Z.
\end{equation}
Any violation of this condition will be due to internal structure. The elementary constituents of a composite particle will naturally conserve their helicity, but they are allowed to jump from one orbital to another, leading thus to a non-conservation of the composite particle's helicity.

The condition \eqref{assumption} expressed in terms of the covariant vertex functions reads
\begin{equation}\label{limint}
\sum_{k=0}^m\frac{\left(-1\right)^{\left[\frac{k+1}{2}\right]}}{2^k}\,C_{[j-k/2]}^{m-k}\,F_{k+1}(0)=\delta_{m,0}\,Z.
\end{equation}
Using the identity
\begin{equation}
\sum_{k=0}^m\left(-1\right)^kC_{[j-k/2]}^{m-k}\,C_{[j+(k-1)/2]}^k=\delta_{m,0},
\end{equation}
we obtain easily the natural values of the covariant vertex functions
\begin{equation}
F_{k+1}(0)=\left(-1\right)^{\left[k/2\right]}2^k\,C_{[j+(k-1)/2]}^k\,Z.
\end{equation}
Multipole form factors, and therefore electromagnetic moments, can be obtained from the covariant vertex functions \cite{partim1}, leading to the simple result
\begin{equation}
\begin{split}
G_{El}(0)+i\,G_{Ml}(0)&=\,i^lC_{2j}^l\,Z,\label{natural}\\
Q_l(0)+\frac{i}{2}\,\mu_l(0)&=\,i^l\frac{(l!)^2}{2^l}\,C_{2j}^l\,Z,
\end{split}
\end{equation}
where $G_{El}$, $G_{Ml}$ are the electric\footnote{If we follow more rigorously the literature terminology, they should be called Coulomb multipoles.} and magnetic multipole form factors, and $Q_l$, $\mu_l$ are the electric and magnetic moments in the natural units $e/M^l$ and $e/2M^l$, respectively.

\subsection{Discussion}

The result we obtain is particularly elegant. It is quite surprising to see that the natural values of the multipole form factors are just given, up to a sign, by a binomial function. These values depend only on the spin $j$ of the particle, the order $l$ of the multipole, and are proportional to the particle electric charge $Z$. Our result also shows explicitly that the highest non-vanishing moment is of order $l=2j$.

Let us develop the expression \eqref{natural} for a particle with unit electric charge $Z=+1$
\begin{displaymath}
\begin{split}
Q_0&=\,G_{E0}(0)=1,\\
\mu_1&=\,G_{M_1}(0)=2j,\\
Q_2&=\,G_{E2}(0)=-j\left(2j-1\right),\\
\mu_3&=\,9\,G_{M3}(0)=-3j\left(2j-1\right)\left(2j-2\right),\\
&\,\,\:\vdots
\end{split}
\end{displaymath}
From the second line, one can see that our assumption about helicity conservation agrees with a universal gyromagnetic factor $g=2$ for any elementary particle \cite{indep}. There the universality of this factor has been proved based on tree-unitarity arguments. Here it arises simply as a consequence of non-trivial light-cone helicity conservation. The general form of the quadrupole (third line) is also in accordance with a good high-energy behavior \cite{higher}. As emphasized in \cite{partim1}, let us also remind that electromagnetic moments and multipole form factors at $Q^2=0$ can only be identified up to order $l=2$.

\begin{table}[h]\begin{center}\caption{\small{The natural multipoles of a spin-$j$ particle with electric charge $Z=+1$ are organized according to a pseudo Pascal triangle, when expressed in terms of natural units of $e/M^l$ and $e/2M^l$ for $G_{El}(0)$ and $G_{Ml}(0)$, respectively.}}\vspace{.5ex}\label{table}
\begin{tabular}{c|ccccccccc}
\hline\hline
$j$&$G_{E0}(0)$&$G_{M1}(0)$&$G_{E2}(0)$&$G_{M3}(0)$&$G_{E4}(0)$&$G_{M5}(0)$&$G_{E6}(0)$&$G_{M7}(0)$&$G_{E8}(0)$\rule{0pt}{3ex}\\
\hline
$0$&$1$&$0$&$0$&$0$&$0$&$0$&$0$&$0$&$0$\rule{0pt}{3ex}\\
$1/2$&$1$&$1$&$0$&$0$&$0$&$0$&$0$&$0$&$0$\rule{0pt}{3ex}\\
$1$&$1$&$2$&$-1$&$0$&$0$&$0$&$0$&$0$&$0$\rule{0pt}{3ex}\\
$3/2$&$1$&$3$&$-3$&$-1$&$0$&$0$&$0$&$0$&$0$\rule{0pt}{3ex}\\
$2$&$1$&$4$&$-6$&$-4$&$1$&$0$&$0$&$0$&$0$\rule{0pt}{3ex}\\
$5/2$&$1$&$5$&$-10$&$-10$&$5$&$1$&$0$&$0$&$0$\rule{0pt}{3ex}\\
$3$&$1$&$6$&$-15$&$-20$&$15$&$6$&$-1$&$0$&$0$\rule{0pt}{3ex}\\
$7/2$&$1$&$7$&$-21$&$-35$&$35$&$21$&$-7$&$-1$&$0$\rule{0pt}{3ex}\\
$4$&$1$&$8$&$-28$&$-56$&$70$&$56$&$-28$&$-8$&$1$\rule[-2ex]{0pt}{5ex}\\
\hline
\end{tabular}\end{center}
\end{table}
The Standard Model (SM) contains only elementary particles up to spin $1$. It requires that at tree level, the elementary weak bosons $W$ and $Z$ have $G_M(0)\equiv G_{M1}(0)=2$ and $G_Q(0)\equiv G_{E2}(0)=-1$, in order to satisfy the Gerasimov-Drell-Hearn sum rule to lowest order in perturbation theory \cite{GDH}. The only known consistent theory for spin-$3/2$ particles is the extended supergravity (SUGRA) \cite{sugra}. The gravitino is described as a spin-$3/2$ particle which couples consistently to electromagnetism, in the framework of $\mathcal{N}=2$ supergravity. One can therefore consider that the multipoles arising from this theory are the natural ones. The result is $G_{E0}(0)=1$, $G_{M1}(0)=3$, $G_{E2}(0)=-3$, and $G_{M3}(0)=-1$ \cite{paper}. In Table \ref{table}, we give the natural values of multipoles we obtain for a particle with unit electric charge $Z=+1$, up to spin $j=4$. As one can see, we are in complete agreement with both SM and SUGRA. The table obtained is just a (pseudo) Pascal triangle because of the binomial function. What is very compelling about our result \eqref{natural} is that it is compact and elegant.

Actually, we noticed that the same result \eqref{natural}, which we derived from the helicity conservation formula \eqref{assumption}, has been obtained in a complementary way in the literature \cite{higherbis}. The authors of Ref. \cite{higherbis} derived model-independent, non-perturbative supersymmetric sum rules for the electromagnetic moments of any theory with $\mathcal{N}=1$ supersymmetry. They find that in any irreducible $\mathcal{N}=1$ supermultiplet, the diagonal matrix elements of the $l^\textrm{th}$ moments are completely fixed in terms of their off-diagonal matrix elements and the diagonal $(l-1)^\textrm{th}$ moments. Setting the off-diagonal matrix elements to zero, any given moment has the same structure for all members of the supermultiplet. This specific case is then considered as leading to the ``preferred'' value of the electromagnetic moments
\begin{equation}
\mathcal{T}_j^{(l)^{(e,m)}}=\mp\frac{1}{M}\,\mathcal{T}_j^{(l-1)^{(m,e)}},
\end{equation}
where the $\mathcal{T}_j^{(l)^{(e,m)}}$ are the generalization\footnote{Rotational invariance allows one to characterize completely each $l^\text{th}$ moment by means of a single quantity. This quantity $\mathcal{T}_j^{(l)^{(e,m)}}$ is essentially a reduced matrix element according to the Wigner-Eckart theorem.} to higher multipoles of the gyromagnetic factor $\mathcal{T}_j^{(1)^{(m)}}\equiv g_j\,\frac{Z\,e}{2M}$. One can easily see that the value of $\mathcal{T}_j^{(l)^{(e,m)}}$ is at the end uniquely fixed by the electric charge $\mathcal{T}_j^{(0)^{(e)}}\equiv Z\,e$. The expression for the $l^\textrm{th}$ moment $M_j^{(l)^{(e,m)}}$ is given by
\begin{equation}\label{SSSR}
M_j^{(l)^{(e,m)}}=\frac{2j\left(2j-1\right)\cdots\left(2j-l+1\right)}{C_{2l}^l}\,\mathcal{T}_j^{(l)^{(e,m)}}.
\end{equation}
To see actually that \eqref{natural} and \eqref{SSSR} do coincide, one has to remember that our moments are given in natural units and take into account that our definition of multipoles \cite{partim1} has an additional factor of $(2l-1)!!$ compared to \cite{higherbis}.

Before closing this section, we would like to comment on a previous study concerning light-cone helicity conservation \cite{Helicity}. The conclusion was that one cannot satisfy at the same time both angular momentum conservation and light-cone helicity conservation. This result is not in contradiction with the present study. In \cite{Helicity}, the author imposes helicity conservation for \emph{any} $Q^2$
\begin{equation}\label{assumption2}
A_{\lambda',\lambda}(Q^2)\propto\delta_{\lambda',\lambda}.
\end{equation}
This assumption is simply different from ours \eqref{assumption}. We impose a \emph{non-trivial} helicity conservation only at the real photon point ($Q^2=0$). Up to spin-$1$, the conclusions are the same. The discrepancies appears when we consider spins higher than $1$. This can be easily understood as follows.

Let us consider the light-cone helicity-flip amplitudes \eqref{LC} for spin $1/2$  
\begin{displaymath}
A_{\frac{1}{2},-\frac{1}{2}}(Q^2)=-\sqrt{\tau}\,F_2(Q^2),
\end{displaymath}
and spin $1$
\begin{displaymath}
\begin{split}
A_{1,0}(Q^2)&=\,\sqrt{2\tau}\left[F_1(Q^2)-\frac{1}{2}\,F_2(Q^2)+\tau\,F_3(Q^3)\right],\\
A_{1,-1}(Q^2)&=\,-\tau\,F_3(Q^2).
\end{split}
\end{displaymath}
All the other helicity-flip amplitudes are related to these ones by discrete space-time symmetries \eqref{parity} and \eqref{time-reversal}. At $Q^2=0$, both assumptions \eqref{assumption} and \eqref{assumption2} about helicity conservation lead to the same result, namely $F_2(0)=0$ for spin $1/2$, and $F_2(0)=2\,F_1(0)$ and $F_3(0)=0$ for spin $1$. Our assumption \eqref{assumption} is less restrictive since only the value at $Q^2=0$ is fixed. The other one \eqref{assumption2} imposes the stronger condition $F_2(Q^2)=2\,F_1(Q^2)$.

Now, for spin $3/2$, let us consider the light-cone helicity-flip amplitudes \eqref{LC}
\begin{displaymath}
\begin{split}
A_{\frac{3}{2},\frac{1}{2}}(Q^2)&=\,\frac{2\sqrt{\tau}}{\sqrt{3}}\left[F_1(Q^2)-\frac{1}{2}\,F_2(Q^2)+\tau\,F_3(Q^3)-\frac{\tau}{2}\,F_4(Q^2)\right],\\
A_{\frac{3}{2},-\frac{1}{2}}(Q^2)&=\,-\frac{2\tau}{\sqrt{3}}\left[F_2(Q^2)+\frac{1}{2}\,F_3(Q^3)+\tau\,F_4(Q^2)\right],\\
A_{\frac{3}{2},-\frac{3}{2}}(Q^2)&=\,\tau^{3/2}\,F_4(Q^3),
\end{split}
\end{displaymath}
and the non-independent one \cite{paper}
\begin{displaymath}
A_{\frac{1}{2},-\frac{1}{2}}(Q^2)=\frac{\sqrt{\tau}}{3}\left[4\,F_1(Q^2)-2\left(1-2\tau\right)F_2(Q^2)+4\tau\,F_3(Q^3)-\tau\left(1-4\tau\right)F_4(Q^2)\right].
\end{displaymath}
Here also, all the other helicity-flip amplitudes are related to these ones by discrete space-time symmetries \eqref{parity} and \eqref{time-reversal}. This set of amplitudes with the requirement of helicity conservation for any $Q^2$ only accepts as a solution the trivial case $F_k(Q^2)=0$ for all $k$. 

Equivalently, one can  obtain the same result by considering angular conditions. While for spin $1$, the unique angular condition \cite{Angcond}
\begin{displaymath}
\left(1+2\tau\right)A_{1,1}(Q^2)-2\sqrt{2\tau}\,A_{1,0}(Q^2)+A_{1,-1}(Q^2)-A_{0,0}(Q^2)=0
\end{displaymath}
is compatible with the requirement of helicity conservation for any $Q^2$, leading to $A_{0,0}(Q^2)=\left(1+2\tau\right)A_{1,1}(Q^2)$, the second of the two angular conditions for spin $3/2$ \cite{paper,Angcond}
\begin{displaymath}
\begin{split}
\left(1+4\tau\right)\sqrt{3}\,A_{\frac{3}{2},\frac{3}{2}}(Q^2)-8\sqrt{\tau}\,A_{\frac{3}{2},\frac{1}{2}}(Q^2)+2\,A_{\frac{3}{2},-\frac{1}{2}}(Q^2)-\sqrt{3}\,A_{\frac{1}{2},\frac{1}{2}}(Q^2)&=\,0,\\
8\tau^{3/2}\,A_{\frac{3}{2},\frac{3}{2}}(Q^2)+2\sqrt{3}\left(1-2\tau\right)A_{\frac{3}{2},\frac{1}{2}}(Q^2)+A_{\frac{3}{2},-\frac{3}{2}}(Q^2)-3\,A_{\frac{1}{2},-\frac{1}{2}}(Q^2)&=\,0,
\end{split}
\end{displaymath}
would imply that $A_{\frac{3}{2},\frac{3}{2}}(Q^2)=0$, \emph{i.e.} no electric charge. So, in general, starting from spin $3/2$ there is at least one angular condition that allows one to write a helicity-conserving amplitude in terms of helicity-flip amplitudes only \cite{Helicity}, leading to a vanishing electric charge under the assumption \eqref{assumption2}. Since with our assumption \eqref{assumption} we found a non-trivial solution for the covariant vertex functions, we are automatically consistent with the angular conditions.

\section{Transverse Spin}\label{TS}

\begin{figure}
	\centering
		\includegraphics[width=8cm]{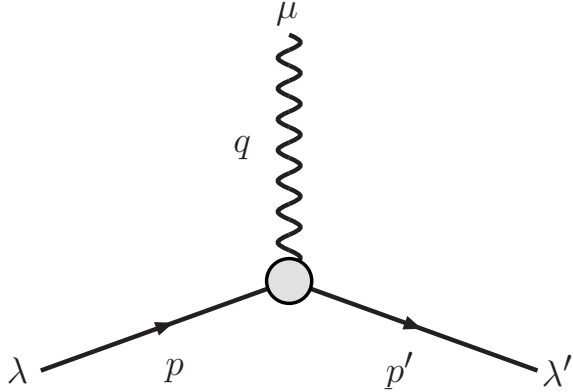}
	\caption{\small{Electromagnetic vertex function (blob). The initial (resp. final) hadron has four-momentum $p$ (resp. $p'$) and helicity $\lambda$ (resp. $\lambda'$). The photon has four-momentum $q=p'-p$ and we work with the light-cone component $\mu=+$.}}\label{fig}
\end{figure}

In this section, we discuss a first application of our helicity conservation assumption \eqref{assumption}. We will consider the electromagnetic $h\to h$ transition (Fig.\ref{fig}) from the viewpoint of a light front moving towards the hadron $h$. This is equivalent to the Infinite Momentum Frame picture where the hadron has a large momentum along the $z$-axis chosen along the direction of $(p'+p)$, where $p$ and $p'$ are as before the initial and final four-momenta of the particle. In the symmetric light-cone frame, the virtual photon couples only to forward-moving partons and the component $J^+_\textrm{EM}(0)$ of the electromagnetic current has the interpretation of quark charge density operator. If one considers only the two light quarks $u$ and $d$, this operator is given by
\begin{displaymath}
J^+_\textrm{EM}(0)=\frac{2}{3}\,\bar u(0)\gamma^+u(0)-\frac{1}{3}\,\bar d(0)\gamma^+d(0).
\end{displaymath}
Each term in this expression is a positive operator since $\bar \psi\,\gamma^+\,\psi\propto|\gamma^+\psi|^2$.

One can define a transverse quark charge density in a hadron with definite light-cone helicity $\lambda$ by the Fourier transform
\begin{equation}
\begin{split}
\rho^h_\lambda(\vec{b})&\equiv\,\int\frac{\ud^2\vec{q}_\perp}{(2\pi)^2}\,e^{-i\vec{q}_\perp\cdot\vec{b}}\,\frac{1}{2p^+}\,\langle
p^+,\frac{\vec{q}_\perp}{2},\lambda'|J^+_\textrm{EM}(0)|p^+,-\frac{\vec{q}_\perp}{2},\lambda\rangle\\
&=\,\int_0^{+\infty}\frac{\ud Q}{2\pi}\,Q\,J_0(Qb)\,A_{\lambda,\lambda}(Q^2).
\end{split}
\end{equation}
For a spin-$j$ hadron $h$, there will be $[j+1]$ independent quark charge densities $\rho^h_\lambda(\vec{b})$ for $\lambda=j,j-1,\cdots,j-[j]$. Quark charge densities provide us therefore with $[j+1]$ independent combinations of covariant vertex functions. To get information for the other covariant vertex functions, we consider also charge densities in a hadron state with the spin transversely polarized.

\subsection{Transverse Spin States}

First, we want to write the transverse spin eigenstates $|j,s_\perp\rangle$ in terms of the helicity eigenstates $|j,\lambda\rangle$. The direction of the transverse polarization is denoted by $\vec{S}_\perp=\cos\phi_S\,\hat e_x+\sin\phi_S\,\hat e_y$. Transverse spin states can be obtained from a rotation of the helicity states
\begin{equation}\label{trans}
|j,s_\perp\rangle=\mathcal{R}(\phi_S,\pi/2,0)\,|j,m=s_\perp\rangle,
\end{equation}
where $\mathcal{R}(\alpha,\beta,\gamma)=e^{-i\alpha J_z}\,e^{-i\beta J_y}\,e^{-i\gamma J_z}$ is the rotation operator with $\alpha,\beta,\gamma$ the Euler angles. Using the Wigner (small) $d$-matrix $d^j_{mm'}(\theta)=\langle j,m|e^{-i\theta J_y}|j,m'\rangle$, we can write \eqref{trans} as
\begin{equation}\label{transverse}
|j,s_\perp\rangle=\sum_{\lambda=j}^j e^{-i\lambda\phi_S}\,d^j_{\lambda s_\perp}(\pi/2)\,|j,\lambda\rangle.
\end{equation}
The explicit expression for the Wigner (small) $d$-matrix is
\begin{displaymath}
\begin{split}
d^j_{mm'}(\theta)=&\,(-1)^{j-m'}\,\sqrt{\left(j+m\right)!\left(j-m\right)!\left(j+m'\right)!\left(j-m'\right)!}\\
&\quad\times \sum_k\frac{(-1)^k\left(\cos\frac{\theta}{2}\right)^{m+m'+2k}\left(\sin\frac{\theta}{2}\right)^{2j-m-m'-2k}}{k!\left(j-m-k\right)!\left(j-m'-k\right)!\left(m+m'+k\right)!},
\end{split}
\end{displaymath}
where the sum is over the integer values of $k$ such that the factorial arguments are non-negative. The properties are
\begin{equation}\label{properties}
\begin{split}
d^j_{mm'}(\theta)&=\,(-1)^{m-m'}\,d^j_{-m-m'}(\theta),\\
&=\,(-1)^{m-m'}\,d^j_{m'm}(\theta),\\
&=\,(-1)^{m-m'}\,d^j_{mm'}(-\theta),\\
&=\,(-1)^{j-m'}\,d^j_{-mm'}(\pi-\theta)\\
\delta_{m,m'}&=\,\sum_{m''=-j}^jd_{m''m}(\theta)\,d_{m''m'}(\theta).
\end{split}
\end{equation}
Since in our case $\theta=\pi/2$, we have
\begin{equation}
d^j_{mm'}(\pi/2)= \sum_k\frac{(-1)^{j-m'+k}\,\sqrt{\left(j+m\right)!\left(j-m\right)!\left(j+m'\right)!\left(j-m'\right)!}}{2^j\,k!\left(j-m-k\right)!\left(j-m'-k\right)!\left(m+m'+k\right)!}.
\end{equation}
Note that for the maximal-spin projection $m'=j$, the expression is very simple $d^j_{mj}(\pi/2)=2^{-j}\,\sqrt{C_{2j}^{j-m}}$.

\subsection{Transversely Polarized Quark Charge Densities}

Next, we want to write the transverse quark charge densities in terms of light-cone helicity amplitudes. The transverse quark charge densities with definite transverse polarization are defined as
\begin{equation}\label{def}
\rho^h_{Ts_\perp}(\vec{b})=\int\frac{\ud^2\vec{q}_\perp}{(2\pi)^2}\,e^{-i\vec{q}_\perp\cdot\vec{b}}\,\rho^h_{Ts_\perp}(\vec{q}_\perp)
\end{equation}
with
\begin{equation}
\rho^h_{Ts_\perp}(\vec{q}_\perp)\equiv \frac{1}{2p^+}\,\langle p^+,\frac{\vec{q}_\perp}{2},s_\perp|J^+(0)|p^+,-\frac{\vec{q}_\perp}{2},s_\perp\rangle.
\end{equation}

Using the change of basis \eqref{transverse}, we can write
\begin{displaymath}
\rho^h_{Ts_\perp}(\vec{q}_\perp)=\sum_{\lambda,\lambda'=-j}^j d^j_{\lambda's_\perp}(\pi/2)\,d^j_{\lambda s_\perp}(\pi/2)\,e^{i\left(\lambda'-\lambda\right)\phi}\,A_{\lambda',\lambda}(Q^2),
\end{displaymath}
with $\phi\equiv\phi_S-\phi_q$. Then, thanks to \eqref{properties}, and both the parity \eqref{parity} and time-reversal \eqref{time-reversal} relations, this can be reduced to
\begin{equation}\label{transdens}
\rho^h_{Ts_\perp}(\vec{q}_\perp)=\sum_{m=0}^{2j}Q^m\,\textrm{Ctrig}\left(m,\phi\right)B_{s_\perp m}(Q^2),
\end{equation}
where we have defined the function
\begin{displaymath}
\textrm{Ctrig}\left(n,\alpha\right)\equiv\frac{1+(-1)^n}{2}\,\cos\left(n\alpha\right)+i\,\frac{1+(-1)^{n+1}}{2}\,\sin\left(n\alpha\right),
\end{displaymath}
and used the compact notation
\begin{equation}\label{transampl}
B_{s_\perp m}(Q^2)=\left(2-\delta_{m,0}\right)\sum_{\lambda'=m/2}^j\left(2-\delta_{\lambda',m/2}\right)d^j_{\lambda's_\perp}(\pi/2)\,d^j_{(\lambda'-m)s_\perp}(\pi/2)\,G_{\lambda',\lambda'-m}(Q^2).
\end{equation}

Finally, using the Bessel function $J_n(x)=\frac{(-i)^n}{2\pi}\int_0^{2\pi}\ud\phi\,\cos\left(n\phi\right)e^{ix\cos\phi}$ in the Fourier transform of \eqref{def} and inserting \eqref{transdens}, we can write the transversely polarized quark charge densities as
\begin{equation}\label{finexpr}
\rho^h_{Ts_\perp}(\vec{b})=\sum_{m=0}^{2j}i^m\,\textrm{Ctrig}\left(m,\phi_b-\phi_S\right)\int\frac{\ud Q}{2\pi}\,Q^{m+1}\,J_m(Qb)\,B_{s_\perp m}(Q^2).
\end{equation}

\subsection{Transverse Electric Moments}

The trigonometric functions appearing in \eqref{finexpr} show that the transversely polarized quark charge densities are not circularly symmetric. Let us therefore consider a bidimensional multipole decomposition by means of circular harmonics. Namely, any function $f(r,\theta)$ with $r\geq 0$ and period $2\pi$ in $\theta$ can be decomposed as
\begin{displaymath}
f(r,\theta)=\sum_{m=-\infty}^{+\infty} f_m(r)\,e^{im\theta}.
\end{displaymath}
Note however that not all circular harmonics are present in \eqref{finexpr}. For a spin-$j$ particle, there are in fact $2j+1$ multipoles. In general, the $l^\textrm{th}$ electric multipole is associated to light-cone amplitudes with $|\lambda'-\lambda|=l$ units of helicity flip. Consequently, it is in principle \emph{sufficient} to know \emph{e.g.} the transversely polarized quark charge density $\rho^h_{Tj}(\vec{b})$ only, in order to have an interpretation of all the $2j+1$ covariant vertex functions.

Choosing the $x$-axis to be parallel to the transverse spin, \emph{i.e.} $\phi_S=0$, we can define the transverse electric moments $Q_{Tl}$ as follows
\begin{equation}
Q_{Tl}\equiv e\int\ud^2\vec{b}\,C_l(\vec{b})\,\rho^h_{Ts_\perp}(\vec{b}),
\end{equation}
where $C_l(\vec{b})$ is given by
\begin{displaymath}
C_l(\vec{b})\equiv b^l\,\textrm{Rtrig}\left(l,\phi_b\right)
\end{displaymath}
with the circular harmonics
\begin{displaymath}
\textrm{Rtrig}\left(n,\alpha\right)\equiv\frac{1+(-1)^n}{2}\,\cos\left(n\alpha\right)+\frac{1+(-1)^{n+1}}{2}\,\sin\left(n\alpha\right).
\end{displaymath}

Working out the Fourier transform leads to
\begin{displaymath}
Q_{Tl}=e\int\ud^2\vec{q}_\perp\,\delta^{(2)}(\vec{q}_\perp)\,C_l(-i\vec{\nabla}_q)\,\rho^h_{Ts_\perp}(\vec{q}_\perp).
\end{displaymath}
Inserting now \eqref{transdens} in this equation gives
\begin{equation}\label{transmoment}
Q_{Tl}=(-1)^{[\frac{l+1}{2}]}\,2^{l-1}\,l!\left(1+\delta_{l,0}\right)\,e\,B_{s_\perp l}(0).
\end{equation}

Let us discuss first the transverse electric charge $Q_{T0}$. A charge is spherically symmetric and should therefore not depend on the orientation of spin. We therefore expect that $Q_{T0}=Q_0$. From \eqref{transmoment} with $l=0$ and \eqref{transampl}, we get
\begin{displaymath}
Q_{T0}=e\sum_{\lambda'=-j}^j \left[d_{\lambda's_\perp}^j(\pi/2)\right]^2\,G_{\lambda',\lambda'}(0).
\end{displaymath}
Thanks to \eqref{consequence} and \eqref{properties}, the expression can be simplified, leading to the expected relation
\begin{displaymath}
Q_{T0}=e\,Z=Q_0.
\end{displaymath}

The other transverse moments are directly proportional to helicity-flip amplitudes. Assuming that helicity at tree level and $Q^2=0$ is non-trivially conserved \eqref{assumption} and using once more Eq. \eqref{consequence}, we conclude that for an elementary particle, the higher transverse electric moments are vanishing. This has as immediate consequence that higher transverse electric moments are just functions of the \emph{anomalous} moments. Such an observation has already been reported for spin up to $3/2$ \cite{spinhalf,spinone,paper}. Our assumption allows us to generalize this observation to any value of the spin.

We would like to emphasize that these higher transverse electric moments are not intrinsic, but are in fact \emph{induced} moments due to a light-cone point of view. We do not claim that particles have \emph{e.g.} an intrinsic dipole electric moment, which would violate parity invariance. This effect is purely induced and is consonant with the observation \cite{Observation} that an object with a magnetic dipole moment at rest, will exhibit an electric dipole moment when moving, orthogonal to both magnetic moment and momentum directions. The magnetic moment of a particle is the source a magnetic dipole field, which is accompanied by an electric field when the particle is moving in a direction different from the magnetic dipole one. Such an electric field will induce electric polarizations in the particle, and thus electric moments, only if the particle has constituents that can migrate. From this point of view, it is clear that the induced polarizations can only be functions of the anomalous moments. 

We can actually use the argument the other way around. Since, on the light-cone, the particle is subject to induced fields that tend to polarize it electrically, its constituents with electric charge (if any) would migrate leading to the appearance of induced electric moments. An elementary particle, \emph{i.e.} structureless or pointlike particle, does not have such constituents and cannot therefore, at tree level, present induced electric moments. As we have shown in this section explicitly, this is equivalent to say that, for an elementary particle, the light-cone helicity amplitudes have to vanish non-trivially at $Q^2=0$ and tree level. The particular values for the usual electromagnetic moments we were able to derive from this condition can therefore be called ``natural''.

\section{Conclusion}

In a set a two papers, we addressed the problem of electromagnetic interaction for arbitrary-spin particles. This problem is an old and a very important one, and requires new constraints in order to be solved. The knowledge of natural electromagnetic moments is rightly one kind of constraints that will help in the construction of a physical Lagrangian theory of electromagnetic interaction with high-spin particles. 

In this paper, we have presented the second part of our results. Using those of Part 1, we have derived the explicit expression of light-cone helicity amplitudes in terms of covariant vertex functions. Under the assumption that light-cone helicity is non-trivially conserved by electromagnetism at $Q^2=0$ and tree level, we have derived the natural value of all electromagnetic moments for any particle. The result turns out to be surprisingly simple. The natural values of multipole form factors at $Q^2=0$ are just given (up to a sign) by a binomial function times the electric charge of the particle.

The result agrees with the values from the Standard Model for elementary spin-$1/2$ (\emph{e.g.} electrons) and spin-$1$ (\emph{e.g.} $W^\pm$ gauge bosons) particles. It is also in accordance with the prediction from $\mathcal{N}=2$ supergravity for gravitinos. Moreover, we also reproduce in a simple way the universality of the gyromagnetic factor $g=2$ and its counterpart for electric quadrupole, as derived from considerations of tree unitarity. Finally, it has also been realized that this result is in fact exactly the one obtained from $\mathcal{N}=1$ supersymmetric sum rules, when one considers that electromagnetic properties do not mix the members of the supermultiplet. All these agreements can hardly be seen as a pure coincidence. Naturally, one still has to completely understand the deep field theoretic implications of non-trivial light-cone helicity conservation. It seems highly probable that this condition can be related to the tree-unitarity argument. This relation is beyond the scope of the present study but will be the object of further investigations.

As an application of our results, we have generalized the discussion about quark transverse charge densities to particles of arbitrary spin. Our assumption concerning helicity conservation directly leads to the conclusion that the transverse higher electric moments can only be functions of the anomalous electromagnetic moments. Using the argument the other way around, an elementary particle cannot present \emph{induced} electric moments in a light-cone framework. This requirement is equivalent to say the light-cone helicity is conserved at tree-level and $Q^2=0$, justifying our assumption \emph{a posteriori}.

\subsection*{Acknowledgements}

The author is grateful to M. Vanderhaeghen, V. Pascalutsa and T. Ledwig for numerous enlightening discussions and comments.

\end{document}